\documentclass{article}
\usepackage{spconf,amsmath,graphicx}
\usepackage{multirow}
\usepackage{amssymb}
\usepackage{cite}
\usepackage{tabularx}
\usepackage{bm,bbm}
\usepackage{makecell}
\usepackage{booktabs}
\usepackage{blindtext}
\usepackage{subfigure}

\usepackage{xurl}
\usepackage{amssymb}
\usepackage[usenames, dvipsnames]{color}

\newcommand{\ve}[1]{\boldsymbol{\mathbf{#1}}}

\title{Contrastive Speaker Embedding With Sequential Disentanglement}
\twoauthors
  {Youzhi~Tu, Man-Wai~Mak\thanks{This work was supported by the RGC of Hong Kong SAR, Grant No. PolyU 15210122.}}
	{Dept. of Electrical and Electronic Engineering\\
	 The Hong Kong Polytechnic University\\
	 Hong Kong SAR}	 
  {Jen-Tzung Chien}
	{Inst. of Electrical and Computer Engineering\\
	 National Yang Ming Chiao Tung University\\
	 Taiwan}

\begin{document}
%
\maketitle
\begin{abstract}
Contrastive speaker embedding assumes that the contrast between the positive and negative pairs of speech segments is attributed to speaker identity only. However, this assumption is incorrect because speech signals contain not only speaker identity but also linguistic content. In this paper, we propose a contrastive learning framework with sequential disentanglement to remove linguistic content by incorporating a disentangled sequential variational autoencoder (DSVAE) into the conventional SimCLR framework. The DSVAE aims to disentangle speaker factors from content factors in an embedding space so that only the speaker factors are used for constructing a contrastive loss objective. Because content factors have been removed from the contrastive learning, the resulting speaker embeddings will be content-invariant. Experimental results on VoxCeleb1-test show that the proposed method consistently outperforms SimCLR. This suggests that applying sequential disentanglement is beneficial to learning speaker-discriminative embeddings.
\end{abstract}

\section{Introduction}
\label{sec:intro}

Modern speaker verification (SV) systems rely on speech data with speaker labels to train an embedding network to achieve state-of-the-art performance. The requirement of speaker labels, however, poses a challenge to system development because manually labeling massive amount of data is expensive and time-consuming. Self-supervised learning has recently emerged as a viable alternative for training speaker embedding networks to circumvent the labeling challenge \cite{Thienpondt20-voxsrc20, Huh20-aat_ssl, Cai21-iter_moco, Zhang21-contrast, Xia21-ssl_moco, Tao22-ssl, Han22-lbl_correct, Cho22-non_contrast, Chen23-dino_reg}.

Self-supervised speaker embedding generally uses contrastive learning \cite{Thienpondt20-voxsrc20, Huh20-aat_ssl, Cai21-iter_moco, Zhang21-contrast, Xia21-ssl_moco, Tao22-ssl} or non-contrastive learning \cite{Han22-lbl_correct, Cho22-non_contrast, Chen23-dino_reg}. The former leads to contrastive embedding, which inevitably faces the class collision issue \cite{Arora19-contrast_theory}, i.e., the negative samples may come from the same speaker as the positive samples in a mini-batch. This problem could drive away the embeddings belonging to the same speaker, resulting in non-discriminative embeddings. Nevertheless, the study in \cite{Zhang21-contrast} has demonstrated that the probability of a mini-batch containing repeated speakers is remarkably low when an appropriate batch size (e.g., 256) is used. This paper focuses on contrastive embedding for text-independent SV.

Contrastive speaker embedding assumes that the contrast between positive and negative pairs is due to speaker identity rather than other explanatory factors of variation \cite{Bengio13-rep_review} like spoken content. However, speaker embeddings contain a variety of information besides speaker identity \cite{Raj19-prob_xvec, Peri20-ana_enc}, and non-speaker factors can also contribute to the contrast between the positive and negative pairs. This erroneous contrast can introduce nuisance information into the embeddings, causing performance degradation. Therefore, it is essential to disentangle the speaker factors from the other factors of variation and only use speaker factors for contrastive learning to ensure that the learned embeddings are \emph{speaker discriminative}. 

In text-independent SV, text content is nuisance and we expect to produce content-invariant speaker embeddings. Because speech signals contain both time-variant and time-invariant information, we can use this inductive bias to factorize the speech representations into static speaker factors and dynamic content factors. For example, in \cite{Li18-dsvae, Bai21-cdsvae}, disentangled sequential variational autoencoders (DSVAEs) were introduced to separate the speaker and content factors in the speaker embeddings. This paper aims to eliminate the effect of content on contrastive learning by incorporating a DSVAE into SimCLR. In particular, a specialized DSVAE is proposed to disentangle the speaker factors from the content factors in the latent space and the resulting speaker factors are used to compute the contrastive loss. In this way, we attribute the contrast between positive and negative pairs to the speakers' identities, producing content-invariant speaker embeddings. 

A recent study with similar motivation as ours is \cite{Lin22-shuffle}. The authors proposed to corrupt the content for each positive pair by shuffling the acoustic frames of one augmented segment. But our proposed method removes content information not only between positive pairs but also between negative pairs. This forces the similarity between the negative pairs to be ascribed to speaker factors, which is amenable to learning speaker-discriminative embeddings. Moreover, instead of shuffling the acoustic frames, our approach removes the dynamic content information in speech frames by separately modeling the content and speaker information using VAEs.


\section{Related Work}
\label{sec:related_works}

Contrastive self-supervised speaker embedding aims to discriminate between the positive and negative pairs of speech segments. Since the emergence of SimCLR \cite{Chen20-simclr} and MoCo \cite{He20-moco}, contrastive speaker embedding has witnessed fast development. For instance, the authors of \cite{Thienpondt20-voxsrc20} applied MoCo to pre-train an embedding network and then used it to generate pseudo labels for iterative refinement. In \cite{Huh20-aat_ssl}, augmentation adversarial training was incorporated into contrastive learning to learn channel-independent embeddings. The authors of \cite{Cai21-iter_moco, Tao22-ssl} used SimCLR to train embedding networks for pseudo-label generation. To address channel variations, the authors of \cite{Zhang21-contrast} proposed channel-invariant training by enforcing the similarity between the clean and augmented embeddings. In \cite{Xia21-ssl_moco}, to alleviate the class collision problem, the authors proposed clustering the whole dataset before negative-segment sampling, so that the negative samples in the prototypical memory bank were more likely from different speakers than the positive sample.

Recently, non-contrastive speaker embedding, especially those based on DINO \cite{Caron21-dino}, has attracted wide attention. In \cite{Cho22-non_contrast, Han22-lbl_correct, Chen23-dino_reg}, the speaker embeddings learned from DINO showed much better performance than the contrastive counterpart. Nevertheless, it has been shown that contrastive and non-contrastive learning objectives are closely related and their optimization is equivalent up to row and column normalization of the embedding matrix \cite{Garrido23-duality}. This discovery suggested that contrastive and non-contrastive methods could perform similarly when the model architectures, loss objectives, and hyperparameters have been sufficiently searched through. This paper focuses on contrastive learning and leaves non-contrastive learning as future work.

\section{Methodology}
\label{sec:methodology}

In this section, we first introduce background knowledge on SimCLR \cite{Chen20-simclr}  and DSVAE \cite{Li18-dsvae, Bai21-cdsvae}. The proposed contrastive learning with sequential disentanglement is then detailed.

\subsection{Contrastive Learning}
\label{subsec:simclr}

SimCLR \cite{Chen20-simclr} is used for contrastive learning in this paper and its schematic is shown in the green block of Fig.~\ref{fig:schematic}. Given a mini-batch of $N$ speech segments $\{ \mathbf{x}_n \}_{n=1}^N$, we obtain $2N$ samples $\{ \tilde{\mathbf{x}}_{n,0}, \tilde{\mathbf{x}}_{n,1} \}_{n=1}^N$ after data augmentation. The segments $\tilde{\mathbf{x}}_{n,0}$ and $\tilde{\mathbf{x}}_{n,1}$ correspond to the same utterance and form a positive pair. For each positive pair $\{ \tilde{\mathbf{x}}_{n,0}, \tilde{\mathbf{x}}_{n,1} \}$, the other $2(N-1)$ augmented segments in the mini-batch are considered negative samples.

Denote the speaker encoder as $f$. We obtain speaker embeddings $\mathbf{e}_{n,i} = f(\tilde{\mathbf{x}}_{n,i})$, where $i \in \{0,1\}$ indexes the augmented segments in a positive pair. For a positive pair of speaker embeddings $\{ \mathbf{e}_{n,i}, \mathbf{e}_{n,|1-i|} \}$, the loss function is defined as 
\begin{equation}
\label{eq:infonce}
\ell_{n,i} = -\log \frac{\exp \left(\operatorname{cos}\left(\mathbf{e}_{n,i}, \mathbf{e}_{n,|1-i|} \right) / \tau\right)} {\sum_{k=1}^{N} \sum_{j=0}^{1} \mathbbm{1}_{[k \neq n,  j \neq i]} \exp \left(\operatorname{cos}\left(\mathbf{e}_{n,i}, \mathbf{e}_{k,j} \right) / \tau\right)},
\end{equation}
where $\mathbbm{1}_{[k \neq n,  j \neq i]}$ denotes an indicator function which is equal to 1 only when $k \neq n$ and $j \neq i$, and $\tau$ is a temperature hyperparameter.
The network is trained by minimizing the \emph{NT-Xent} loss \cite{Chen20-simclr}:
\begin{equation}
\label{eq:nt-xent}
\mathcal{L}_{\mathrm{C}}=\frac{1}{2N} \sum_{n=1}^N \left[\ell_{n,0}+\ell_{n,1}\right].
\end{equation}

\subsection{Disentangled Sequential Variational Autoencoder (DSVAE)}
\label{subsec:amsoftmax}

As shown in the orange blocks of Fig.~\ref{fig:schematic}, a DSVAE  \cite{Li18-dsvae} consists of a speaker encoder $f^\mathrm{s}$, a content encoder $f^\mathrm{c}$, and a decoder $g$.\footnote{To better fit into the context of speaker verification, our terminologies differ from those in [14]. Specifically, our content means linguistic content, whereas the content in [14] means static content, e.g., object identity.}  Consider a sequence of filter-bank features with $T$ frames $\mathbf{x}_{1: T}=\left(\mathbf{x}_1, \ldots, \mathbf{x}_T\right)$, two groups of latent variables (embeddings) $\mathbf{e}^\mathrm{s}$ and $\mathbf{e}^\mathrm{c}_{1: T}$ are encoded in the latent space. The sequence generation process can be formulated as
\begin{align}
\label{eq:dsvae_generation}
p\left( \mathbf{x}_{1: T}, \mathbf{e}^\mathrm{s}, \mathbf{e}^\mathrm{c}_{1: T} \right)
&= p(\mathbf{e}^\mathrm{s}, \mathbf{e}^\mathrm{c}_{1: T}) 
   p\left(\mathbf{x}_{1: T} | \mathbf{e}^\mathrm{s}, \mathbf{e}^\mathrm{c}_{1: T}\right) \nonumber \\ 
&= p(\mathbf{e}^\mathrm{s}) 
   \prod_{t=1}^T p\left( \mathbf{e}^\mathrm{c}_t | \mathbf{e}^\mathrm{c}_{<t} \right)
   p\left( \mathbf{x}_t | \mathbf{e}^\mathrm{s}, \mathbf{e}^\mathrm{c}_t  \right),
\end{align}
where $\mathbf{e}^\mathrm{c}_{<t}\equiv(\mathbf{e}^\mathrm{c}_{0}, \ldots, \mathbf{e}^\mathrm{c}_{t-1})$ and $\mathbf{e}^\mathrm{c}_{0}=\mathbf{0}$. 

To approximate the posterior of the latent variables, we use a variational inference model
\begin{align}
\label{eq:dsvae_inference}
q\left( \mathbf{e}^\mathrm{s}, \mathbf{e}^\mathrm{c}_{1: T} | \mathbf{x}_{1: T}\right)
&= q\left(\mathbf{e}^\mathrm{s} | \mathbf{x}_{1: T}\right) 
   q\left(\mathbf{e}^\mathrm{c}_{1: T} | \mathbf{x}_{1: T}\right) \nonumber \\ 
&= q\left(\mathbf{e}^\mathrm{s} | \mathbf{x}_{1: T}\right) 
   \prod_{t=1}^T q\left(\mathbf{e}^\mathrm{c}_t | \mathbf{e}^\mathrm{c}_{<t}, \mathbf{x}_{\leq t} \right).
\end{align}
The speaker latent posterior follows a Gaussian distribution $q\left(\mathbf{e}^\mathrm{s} | \mathbf{x}_{1: T}\right) = \mathcal{N} \left( \mathbf{e}^\mathrm{s} ; \ve{\mu}^\mathrm{s}(\mathbf{x}_{1: T}), \mathrm{diag}\{(\ve{\sigma}^\mathrm{s}(\mathbf{x}_{1: T}))^2\} \right)$, where $\ve{\mu}^\mathrm{s}(\cdot)$ and $\ve{\sigma}^\mathrm{s}(\cdot)$ are modeled by a speaker embedding network with two linear heads. Similarly, we have $q\left(\mathbf{e}^\mathrm{c}_t | \mathbf{e}^\mathrm{c}_{<t}, \mathbf{x}_{\leq t} \right) = \mathcal{N} \left( \mathbf{e}^\mathrm{c}_t ; \ve{\mu}^\mathrm{c}_t(\mathbf{e}^\mathrm{c}_{<t}, \mathbf{x}_{\leq t}) , \mathrm{diag} \{(\ve{\sigma}^\mathrm{c}_t(\mathbf{e}^\mathrm{c}_{<t}, \mathbf{x}_{\leq t}))^2\} \right)$, and $\ve{\mu}^\mathrm{c}_t(\cdot)$ and $\ve{\sigma}^\mathrm{c}_t(\cdot)$ can be modeled by LSTMs \cite{Hochreiter97-lstm}. To sample $\mathbf{e}^\mathrm{s}$ and $\mathbf{e}^\mathrm{c}_{1: T}$, the reparameterization trick \cite{Kingma14-vae} is used.
  
We define the DSVAE loss as the nagative of the evidence lower bound (ELBO) \cite{Bai21-cdsvae} of log likelihood as follows
{\footnotesize
\begin{align}
\label{eq:dsvae_loss}
\mathcal{L}_{\mathrm{DSVAE}} &= -\mathbb{E}_{p_{\scalebox{0.5}{{D}}} (\mathbf{x}_{1: T})} \mathbb{E}_{q\left(\mathbf{e}^\mathrm{s}, \mathbf{e}^\mathrm{c}_{1: T} | \mathbf{x}_{1: T}\right)} \Bigl[ \log p\left(\mathbf{x}_{1: T} | \mathbf{e}^\mathrm{s}, \mathbf{e}^\mathrm{c}_{1: T} \right) \Bigr] \nonumber \\
& \quad + \mathrm{KL} \Bigl[ q\left(\mathbf{e}^\mathrm{s} | \mathbf{x}_{1: T}\right) \| p(\mathbf{e}^\mathrm{s}) \Bigr] 
+ \mathrm{KL} \Bigl[ q\left(\mathbf{e}^\mathrm{c}_{1: T} | \mathbf{x}_{1: T}\right) \| p\left(\mathbf{e}^\mathrm{c}_{1: T}\right) \Bigr] \nonumber \\
& \quad - \left[I\left(\mathbf{e}^\mathrm{s} ; \mathbf{x}_{1: T}\right) + I\left(\mathbf{e}^\mathrm{c}_{1: T} ; \mathbf{x}_{1: T}\right)\right]
+ I\left(\mathbf{e}^\mathrm{s}; \mathbf{e}^\mathrm{c}_{1: T} \right),
\end{align}}where $p_{\scalebox{0.5}{{D}}}(\mathbf{x}_{1: T})$ is the empirical distribution of the sequence, $\mathrm{KL}[\cdot\|\cdot]$ denotes Kullback-Leibler (KL) divergence, and $I(\cdot;\cdot)$ is mutual information (MI). The first term of Eq.~\ref{eq:dsvae_loss} represents the reconstruction loss and the subsequent two terms are KL divergence between the posteriors and the priors w.r.t. $\mathbf{e}^\mathrm{s}$ and $\mathbf{e}^\mathrm{c}_{1: T}$, respectively. The maximization of the MI between the latent variables and the input preserves respective information in the embeddings, whereas minimizing the MI between $\mathbf{e}^\mathrm{s}$ and $\mathbf{e}^\mathrm{c}_{1: T}$ encourages their independence, resulting in disentangled embeddings $\mathbf{e}^\mathrm{s}$. In practice, a kind of information bottleneck features is implemented.

\subsection{Contrastive Speaker Embedding  With Sequential Disentanglement}
\label{subsec:ds_emb}

\begin{figure}[t]
\centering
\includegraphics[width=.5\textwidth]{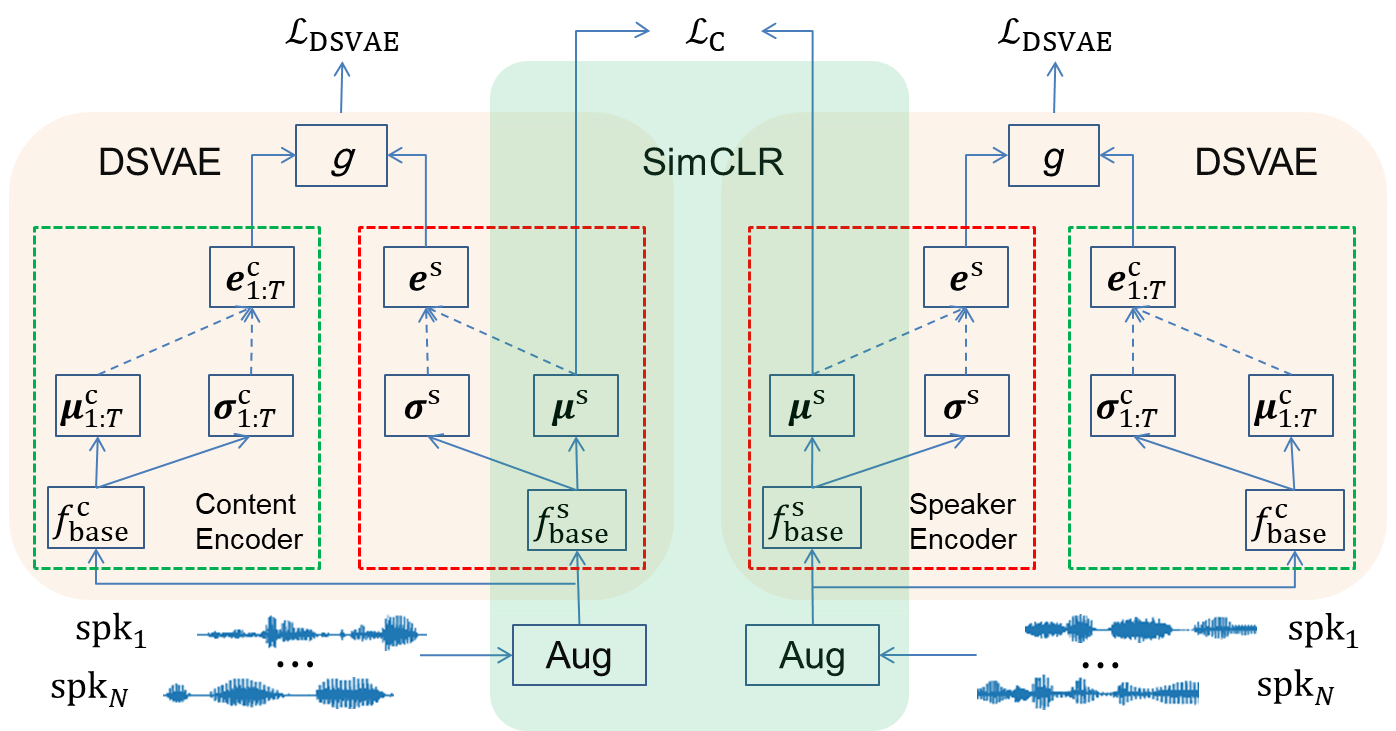}
\caption{Schematic of contrastive speaker embedding with sequential disentanglement. This method incorporates a DSVAE (orange block) into SimCLR (green block). The dashed arrows within the speaker encoder and the content encoder denote Gaussian sampling. After training, the vectors produced by the $\ve{\mu}^\mathrm{s}$ node are used as speaker embeddings.}
\label{fig:schematic}
\end{figure}

In conventional contrastive speaker embedding, the contrast between positive and negative pairs can be attributed to speaker identity and content. To learn content-invariant embeddings, we propose using a DSVAE to disentangle the speaker factors from the content factors and remove the content factors from contrastive learning.

The schematic of the proposed method is shown in Fig.~\ref{fig:schematic}. The base speaker encoder $f^\mathrm{s}_\mathrm{base}$, the mean head $\ve{\mu}^\mathrm{s}$, the standard deviation head $\ve{\sigma}^\mathrm{s}$, and the Gaussian sample head $\ve{e}^\mathrm{s}$ constitute the speaker encoder $f^\mathrm{s}$. The same applies to the content encoder $f^\mathrm{c}$. Note that $f^\mathrm{s}_\mathrm{base}$ and $f^\mathrm{c}_\mathrm{base}$ can share the lower frame-level layers. 

To train the network, we define the total loss:
\begin{equation}
\label{eq:total_loss}
\mathcal{L}=\mathcal{L}_{\mathrm{C}} + \lambda \mathcal{L}_{\mathrm{DSVAE}},
\end{equation}
where $\lambda$ controls the contribution of sequential disentanglement. Note that when computing $\mathcal{L}_{\mathrm{C}}$, the embeddings $\mathbf{e}_{n,i}$ in Eq.~\ref{eq:infonce} should be replaced by $\ve{\mu}^\mathrm{s}_{n,i}$ because the content variables have been discarded. After successful training, we extract the vectors from the $\ve{\mu}^\mathrm{s}$ node as speaker embeddings.

\section{Experimental Setup}
\label{sec:exp}

In this paper, SimCLR and the proposed method were evaluated on the original (Vox1-O), extended (Vox1-E), and hard (Vox1-H) VoxCeleb1 test sets \cite{Nagrani20-voxceleb}. The VoxCeleb2 development set was used for training the embedding networks, which comprised 1,092,009 utterances from 5,994 speakers.

\subsection{Acoustic Feature Extraction}
\label{subsec:exp_acoustic}

We extracted 80-dimensional filter-bank features from each utterance with a 25-ms window and a 10-ms frame shift. Cepstral mean normalization was applied to the extracted features. Before acoustic feature extraction, data augmentation was performed by adding reverberation, noise, music, and babble to the original speech signals. The MUSAN \cite{Snyder15-musan} corpus was used as the additive noise sources. For creating reverberated speech signals, we convolved the original speech signal with the simulated room impulse responses (RIRs).\footnote{https://www.openslr.org/28/rirs\_noises.zip.}

\begin{table*}[t] 
  \caption{Performance on VoxCeleb1 test sets. The results in Rows 4--7 were obtained without iterative model refinement. The minDCF with a superscript $*$ in Rows 4--5 was calculated using $P_{\mathrm{target}}=0.05$ instead of $P_{\mathrm{target}}=0.01$.}
  \label{tab:main_res}
  \centering
  \scalebox{0.72}{
	\begin{tabular}{c|c|cc|cc|cc}
		\toprule
		& & \multicolumn{2}{c|} {Vox1-O} & \multicolumn{2}{c|} {Vox1-E} 
		& \multicolumn{2}{c} {Vox1-H} \\
		Row & Contrastive Learning Framework & EER (\%) & minDCF & EER (\%) & minDCF 
		& EER (\%) & minDCF \\
		\midrule
		\midrule
		1 & SimCLR & 7.13 & 0.571 & 7.89 & 0.596 & 12.26 & 0.692  \\
		2 & SimCLR + Frame Shuffle \cite{Lin22-shuffle} 
		  & 7.90 & 0.570  & 8.48 & 0.600 & 12.86 & 0.737  \\ 
		3 & SimCLR + DSVAE (Proposed) 
		  & \bf 6.37 & \bf 0.533 & \bf 7.36 & \bf 0.574 & \bf 11.72 & \bf 0.677 \\
		\midrule 
		4 & \cite{Huh20-aat_ssl}, added augmentation adversarial training to SimCLR
		  & 8.65 & 0.454$^*$ & -- & -- & -- & --  \\
		5 & \cite{Cai21-iter_moco}, used SimCLR with adapted contrastive loss objective
		  & 8.86 & 0.508$^*$ & 10.15 & 0.570$^*$ & 16.20 & 0.710$^*$  \\
		6 & \cite{Zhang21-contrast}, added channel-invariant training to SimCLR
		   & 8.28 & 0.610 & -- & -- & -- & --  \\
		7 & \cite{Tao22-ssl}, added augmentation adversarial training to SimCLR
		   & \bf 7.36 & -- & \bf 7.90 & -- & \bf 12.32 & --  \\
		\bottomrule
	\end{tabular}
	}
\end{table*}

\subsection{Network Structure and Training}
\label{subsec:exp_network}

A 512-channel ECAPA-TDNN \cite{Desplanques20-ecapatdnn} was used as the base encoder. On top of the pooling layer, three linear heads of 192 nodes each were added for $\ve{\mu}^\mathrm{s}$, $\ve{\sigma}^\mathrm{s}$, and $\ve{e}^\mathrm{s}$, respectively. The lowest four layers of the content and speaker encoders were shared. On top of the shared layers, we sequentially added a bidirectional LSTM layer of 512 hidden nodes and a 512-node RNN layer to the content encoder. Three linear layers of 32 hidden nodes were added to represent $\ve{\mu}^\mathrm{c}_t$, $\ve{\sigma}^\mathrm{c}_t$, and $\ve{e}^\mathrm{c}_t$, respectively. The decoder comprised two convolutional layers with 512 and 80 channels, respectively. The dilation rates of these two layers were 2 and 1, respectively.

We used Adam for optimization. A linear learning rate warmup was employed during the first 10 epochs, increasing the learning rate from 1e-4 to 1e-3. After that, it was decayed to 1e-5 with a cosine scheduler. Totally, the networks were trained for 50 epochs. The mini-batch size was set to 256. Each mini-batch was created by randomly selecting speech segments of 3.5 seconds from the training set and the two segments of each positive pair can overlap. The temperature in  Eq.~\ref{eq:infonce} was set to 0.05 and we used 0.01 for the $\lambda$ in Eq.~\ref{eq:total_loss}.

\section{Results and Discussions}
\label{sec:res}

Cosine similarity was used in all experiments. The performance was evaluated w.r.t. equal error rate (EER) and minimum detection cost function (minDCF) at  $P_{\mathrm{target}}=0.01$.

\subsection{Performance on VoxCeleb1 Test}
\label{subsec:main_res}

The results on VoxCeleb1-test are shown in Table~\ref{tab:main_res}. Rows 4--7 show the results of existing SimCLR without iterative model refinement, and Row 1 shows that our baseline system performs better. From Rows 1 and 3, we observe that the proposed method consistently outperforms SimCLR across three tasks. This indicates that it is effective to apply a DSVAE to disentangle the speaker factors from the content factors and only use speaker factors for contrastive learning.

We also employed frame shuffling \cite{Lin22-shuffle} for comparison. As shown in Rows 1--2, shuffling the frames degrades the performance  compared with the standard SimCLR. This observation contradicts with that in \cite{Lin22-shuffle}. One reason of performance degradation could be that shuffling the frames corrupts the underlying acoustic dynamics that are beneficial to both speech and speaker recognition. Although the content dependency is destroyed as expected, the information related to speaker discrimination is also undermined. An empirical evidence can be found in \cite{Li22-feat_shuffle}, where the authors observed that shuffling the acoustic sequence at the frame level degrades SV performance. This is in accordance with the results in Table~\ref{tab:main_res}.

\begin{table}[t] 
  \caption{Performance of different representations on Vox1-O. ``spk\_emb" denotes speaker embeddings ($\mathbf{e}^\mathrm{s}$ in Eq.~\ref{eq:dsvae_generation}). ``avg\_con\_emb" means temporally-averaged content embeddings ($\mathbf{e}^\mathrm{c}_{1: T}$ in Eq.~\ref{eq:dsvae_generation}). ``Init" in Row 5 means that the model is randomly initialized without training.}
  \label{tab:spk_info}
  \centering
  \scalebox{0.72}{
	\begin{tabular}{c|c|c|cc}
		\toprule
		& & & \multicolumn{2}{c} {Vox1-O}  \\
		Row & Framework & Representation & EER (\%) & minDCF  \\
		\midrule
		\midrule
		1 & SimCLR & spk\_emb & 7.13 & 0.571  \\
		2 & DSVAE & spk\_emb & 22.87 & 0.915  \\
		3 & SimCLR + DSVAE & spk\_emb & \bf 6.37 & \bf 0.533   \\
		4 & SimCLR + DSVAE & avg\_con\_emb & 47.51 & 0.998 \\
		5 & SimCLR + DSVAE (Init) & avg\_con\_emb & 41.32 & 0.994   \\
		\bottomrule
	\end{tabular}
  }
\end{table}

To further verify the benefit of sequential disentanglement, we investigate the speaker information in the speaker embeddings ($\mathbf{e}^\mathrm{s}$ in Eq.~\ref{eq:dsvae_generation}) and content embeddings ($\mathbf{e}^\mathrm{c}_{1: T}$ in Eq.~\ref{eq:dsvae_generation}). Table~\ref{tab:spk_info} shows the performance of different representations on Vox1-O. From Row 2, we observe that the performance of the speaker embeddings extracted from DSVAE is much worse than that of SimCLR. This suggests that using DSVAE alone is not sufficient to separate the speaker and content information. Row 4 and Row 5 show the performance of the proposed framework with and without training, respectively. We see that there is almost no speaker information in the temporally-averaged content embeddings after random initialization and disentangled learning can further reduce the speaker information in the content embeddings. In short, combining DSVAE and SimCLR is effective to learn discriminative speaker embeddings.

\subsection{Effect of $\lambda$}
\label{subsec:res_lambda}

\begin{figure}[t]
\centering
\subfigure[]{
\includegraphics[width=.22\textwidth]{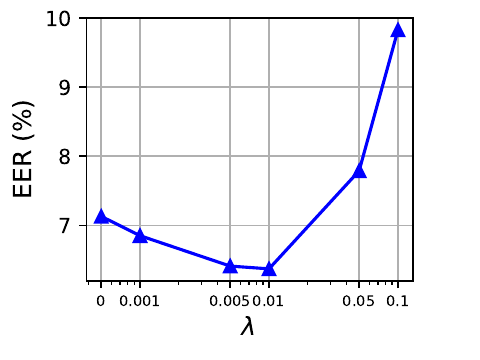}
}
\subfigure[]{
\includegraphics[width=.22\textwidth]{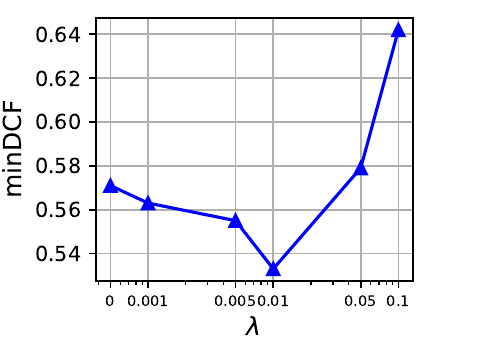}
}
\caption{Comparison of (a) EER and (b) minDCF of the proposed method with respect to $\lambda$ on Vox1-O.}
\label{fig:res_lambda}
\end{figure}

The hyperparameter $\lambda$ in Eq.~\ref{eq:total_loss} controls the contribution of DSVAE to the proposed method. We investigate the effect of $\lambda$ and the results are shown in Fig.~\ref{fig:res_lambda}. The best result was obtained at $\lambda=0.01$ for both EER and minDCF. When $\lambda$ is larger than 0.05, the performance begins to degrade. This shows that too much emphasis on sequential disentanglement can prevent the model from learning discriminative embeddings, which complies to the result in Row 2 of Table~\ref{tab:spk_info}.

\section{Conclusions}
\label{sec:conclusions}

This paper proposed a contrastive learning framework with sequential disentanglement for text-independent SV. The proposed method adopted a DSVAE for disentangling speaker factors from content factors and used the speaker factors only for contrastive learning. Results on VoxCeleb1 test sets showed that the proposed method consistently outperforms SimCLR, suggesting that it is beneficial to incorporate sequential disentanglement into contrastive learning for learning speaker-discriminative embeddings.

\vfill\pagebreak

{\small
\bibliographystyle{IEEEbib}
\bibliography{mybib}
}

\end{document}